\documentclass[12pt]{article}
\usepackage{times}
\usepackage{geometry}
\usepackage[utf8]{inputenc}
\usepackage{enumitem,amssymb}
\usepackage{ragged2e}
\newlist{thematic}{itemize}{8}
\setlist[thematic]{label=$\square$}
\usepackage{pifont}
\usepackage{color,graphicx,hanging}
\newcommand{\xmark}{\ding{55}}%

\newcommand{\kms}{\mbox{km~s$^{-1}$}}
\newcommand{\juc}{\mbox{$J$=1$-$0}}
\def\lsim{~\rlap{$<$}{\lower 1.0ex\hbox{$\sim$}}}
\def\gsim{~\rlap{$>$}{\lower 1.0ex\hbox{$\sim$}}}
\newcommand{\jdu}{\mbox{$J$=2$-$1}}

\begin{document}
\raggedright
\huge
Astro2020 Science White Paper \linebreak

Molecular Masers as Probes of the Dynamic Atmospheres of Dying Stars\linebreak
\normalsize

\noindent \textbf{Thematic Areas:} \hspace*{60pt} $\square$ Planetary Systems \hspace*{10pt} $\square$ Star and Planet Formation \hspace*{20pt}\linebreak
$\square$ Formation and Evolution of Compact Objects \hspace*{31pt} $\square$ Cosmology and Fundamental Physics \linebreak
  $\square$\hspace{-8pt}\xmark\hspace{1pt}  Stars and Stellar Evolution \hspace*{1pt} $\square$ Resolved Stellar Populations and their Environments \hspace*{40pt} \linebreak
  $\square$    Galaxy Evolution   \hspace*{45pt} $\square$             Multi-Messenger Astronomy and Astrophysics \hspace*{65pt} \linebreak
  
\textbf{Principal Author:}

Name:	Lynn D. Matthews
 \linebreak						
Institution:  Massachusetts Institute of Technology Haystack Observatory
 \linebreak
Email: lmatthew@mit.edu
 \linebreak
%Phone:  1-617-715-5400
% \linebreak
 
\textbf{Co-authors:}\\ Mark J Claussen (National Radio Astronomy Observatory)
 \linebreak Graham M. Harper (University of Colorado - Boulder) \linebreak

\textbf{Executive Summary:}
More than half of the dust and heavy element enrichment in galaxies
originates from the winds and outflows of evolved, low-to-intermediate
mass stars on the asymptotic giant branch (AGB). However, numerous 
details of the physics of late-stage stellar
mass loss remain poorly understood, ranging from the wind launching
mechanism(s) to the geometry and timescales of the mass loss. One of the major
challenges to understanding AGB winds is that the AGB
evolutionary phase is
characterized by the interplay between highly complex and dynamic processes, including radial
pulsations, shocks, magnetic fields, opacity changes
due to dust and molecule formation, and large-scale convective
flows. Collectively, these phenomena lead to changes in the observed stellar
properties on timescales of days to years. Probing the complex
atmospheric physics of AGB stars therefore demands exquisite
spatial resolution, coupled with temporal monitoring over both short and long
timescales. Observations of the molecular maser lines that arise in the winds and
outflows of AGB stars using very long baseline interferometry (VLBI) offer one of the most powerful tools available
to measure the atmospheric dynamics, physical conditions, and magnetic
fields with ultra-high spatial
resolution (i.e.,  tens of $\mu$s, corresponding to $\sim0.002R_{\star}$ at
$d\approx$150~pc), coupled with the ability to track features and phenomena on
timescales of days to years. 
Observational advances in the coming decade will enable
contemporaneous observations of an unprecedented number of maser transitions spanning
centimeter to submillimeter wavelengths. In evolved stars,
observations of masers within the winds and outflows 
are poised to provide groundbreaking new insights into the atmospheric
physics and mass-loss process.

\begin{justify}
\bigskip\bigskip {\it Related White Papers:} 
L. D. Matthews et al., Unlocking the Secrets of Late-Stage Stellar Evolution and Mass Loss through Radio
Wavelength Imaging

\pagebreak
%Insert your white paper text here (max of five pages including figures).

\section{Context\protect\label{intro}}
\vspace{-0.3cm}
For low-to-intermediate mass stars  ($0.8 \lsim M_{*} \lsim
8~M_{\odot}$), the end stages of stellar
evolution are marked by the onset of a range of complex and dynamic
atmospheric phenomena. During the asymptotic giant branch (AGB) phase,
such stars exhibit dramatic increases  in radius ($R_{\star}\gsim$1--2~AU) and
luminosity ($L_{\star}\sim10^{4}L_{\odot}$) and begin undergoing
radial pulsations with periods of order hundreds of days. Over the
course of single pulsation cycle 
the brightness of the star may vary by up to $\sim$~8 magnitudes (a
factor of 1000) in
the visible. The latter variations are attributed to the time-dependent formation
of metallic oxides in the outer atmosphere (Reid \& Goldston 2002).

Another key hallmark of the AGB phase is the onset of periods of intense mass
loss   (${\dot M}\sim10^{-8}$--$10^{-4}~M_{\odot}$
yr$^{-1}$) through cool, dense, low-velocity
winds   ($V_{\rm out}\sim$10~\kms). These outflows ultimately expel up to
80\% of the star's initial mass (see review by H\"ofner \& Olofsson
2018), leading to profound effects on the stellar evolutionary
track. Because AGB winds
are dusty and enriched in heavy elements, AGB mass loss  also 
produces more than half of the  dust and chemical enrichment in the Galaxy, (Schr\"oder \&
Sedlmayr 2001; Van Eck et al. 2001; Karakas 2014). 

A detailed understanding of AGB mass loss is crucial 
for stellar astrophysics and knowledge of the ultimate fate of our
Sun. But more broadly, AGB stars impact the entire Galactic ecosystem,
and prescriptions for the mass loss and dust and heavy element
production are crucial  for
extragalactic astronomy and cosmology, which make use of
stellar population synthesis  (e.g., Salaris et al. 2014),
interpretations of the integrated starlight from galaxies
(e.g., Melnick \& De Propris 2013), and  prescriptions of gas recycling and
chemical evolution in galaxies (e.g., Tosi 2007; Leitner \& Kravtsov 2011).
However, we still lack a comprehensive and
self-consistent picture of evolution and mass loss along the
AGB. Persistent uncertainties  include the wind launching
mechanisms for  stars of different chemistries, the mass-loss
geometry and timescales, and the evolutionary pathways
for stars of various initial masses
(Marengo 2009; H\"ofner \& Olofsson 2018).

In broad terms,
AGB winds are thought to be dust-driven   (e.g., Kwok 1975): dust
formation occurs in the cool, outer atmosphere ($r>2R_{\star}$) and
radiation pressure on these grains transfers momentum outward to the
gas, leading
to mass loss. However, in  warmer and/or oxygen-rich (M-type) stars,
conditions are too hot for dust formation interior to
$r\sim$6--7~AU. Thus some additional mechanism is required to
transport material from the stellar ``surface'' to the wind launch
region  (Woitke 2006; H\"ofner 2011).
Pulsations are suspected of playing
a critical role in this process (Bowen 1988; Yoon \& Cantiello 2010;
Neilson 2014), but the details are still poorly understood. Magnetic fields, acoustic waves, and/or Alfv\'en waves
are also candidates for shaping and regulating mass loss (e.g., Blackman et
al. 2001; Harper 2010), possibly in conjunction
with large-scale convective processes (e.g., Lim et al. 1998; O'Gorman et
al. 2017). However, the intricate interplay between these various processes
is still poorly constrained.
Significant progress in this field will require empirical
constraints that combine:  (1) the ability to {\em
  spatially resolve} the stellar atmosphere on the relevant physical scales ($\ll
R_{\star}$); (2)  {\it temporal resolution} of the characteristic 
dynamical timescales; and (3) the ability to {\em directly measure
gas motions}. 

Fortuitously, the molecule-rich atmospheres of evolved giants
frequently give rise to {\em molecular masers}  that can be exploited
for this purpose (e.g., Gray 2012).  Though stellar masers
were discovered $\sim$50 years ago (Wilson \& Barrett 1968; Snyder \& Buhl 1974),
technological advances, coupled with advances in
maser theory (Gray et al. 2016) and the modeling of AGB star atmospheres
(Freytag et al. 2017; H\"ofner \& Freytag 2019), are
poised to allow masers to supply tremendous new insights into the physics of evolved stars
and stellar mass loss in the coming decade. 

\section{Molecular Masers in Evolved Stellar Atmospheres: Background and
  Recent Results}
\vspace{-0.2cm}
In oxygen-rich (M-type) AGB stars, maser emission from SiO
arises within a few $R_{\star}$, just
outside the radio photosphere ($r\sim2R_{\star}$), and adjacent to the
dust formation zone and molecular layers (Reid \& Menten 1997; Fig.~\ref{fig:cross-section}).    
The properties of the SiO masers are therefore intricately linked
with the atmospheric regions where stellar mass 
 loss originates   (Humphreys 2002; Gray et
al. 2009). In carbon-rich AGB stars, HCN masers are thought to trace similar
regions (Gray 2012; Izumiura et al. 1995; Menten
et al. 2018).
H$_{2}$O masers, in comparison, typically arise just outside the dust
formation zone at $r\sim10$--100~AU.   

%%%%%%%%%%%%%%%%%%%%%%%%%%%%%%%%%%%%%%%%
\begin{figure}[!t]
\vspace{-0.1cm}
\centering
\scalebox{0.4}{\rotatebox{0}{\includegraphics{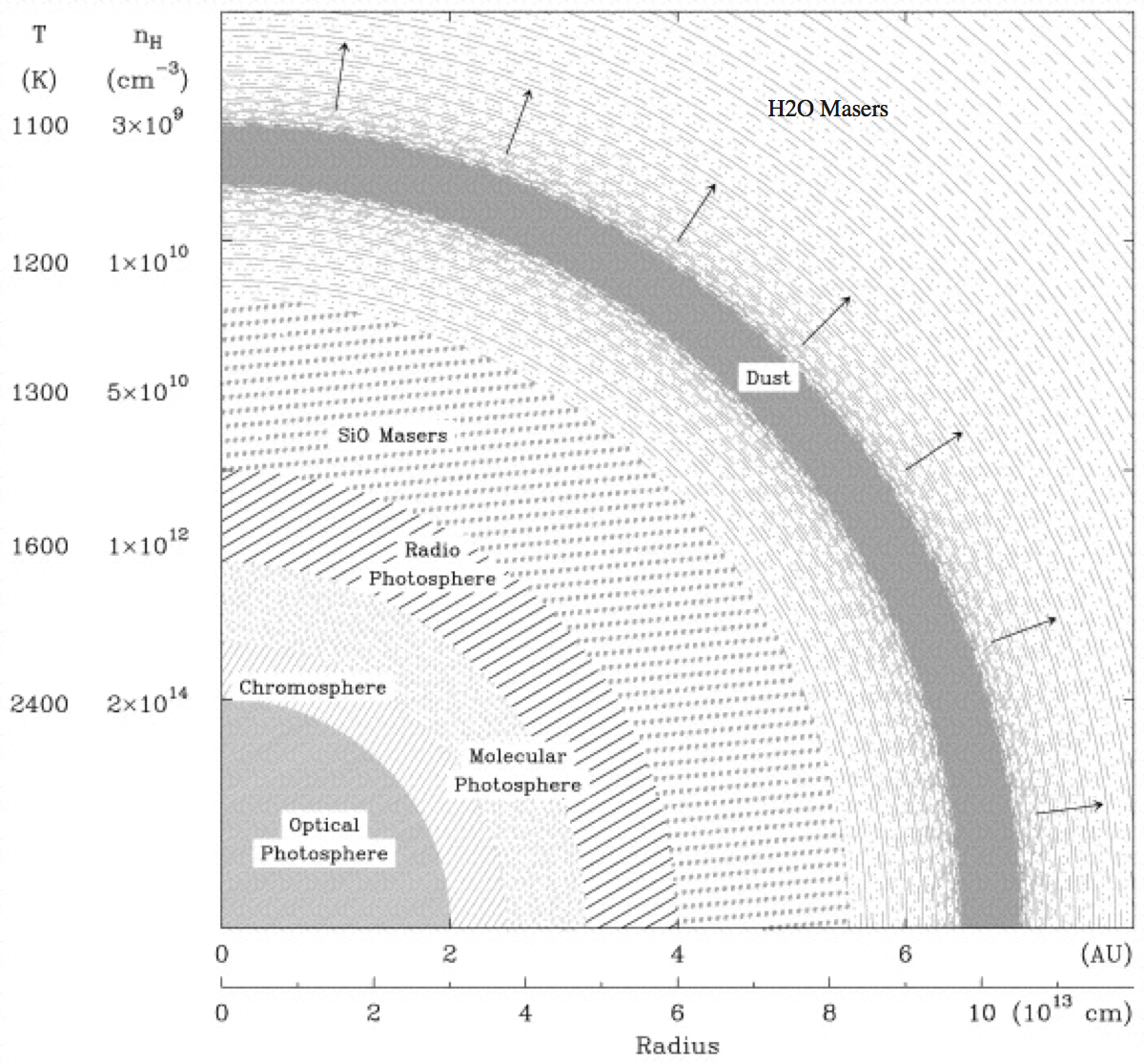}}}
\caption{{\small Schematic illustrating the various atmospheric layers
    of an M-type AGB star. SiO maser emission arises near the
    radio
    photosphere, just interior to the
    dust formation zone where the stellar wind is launched. H$_{2}$O
    masers arise just outside the dust formation zone. Adapted from Reid \& Menten (1997).}
\protect\label{fig:cross-section}}
\vspace{-0.4cm}
\end{figure}
%%%%%%%%%%%%%%%%%%%%%%%%%%%%%%%%%%%%%%%

Because of their high brightness temperatures ($\gsim10^{6}$~K), masers
can be observed with ultra-high spatial resolution using very long baseline
interferometry (VLBI). In recent years, VLBI observations with the Very
Long Baseline Array (VLBA) of stellar SiO and H$_{2}$O masers
with angular resolutions of $\sim$0.2--0.5~mas have
established the enormous potential of high-resolution studies
of masers for understanding the complex atmospheric physics
and mass loss of AGB stars. Examples of key results to date include:

\vspace{-0.4cm}
\paragraph{Spatial structure}
Observations with the VLBA have
revealed that SiO masers lie in complex ring-like structures centered
on the host star, lying just outside
the hot molecular layer observable at IR wavelengths  
($r\sim$2--4$R_{\star}$; e.g., Diamond et al. 1994; Cotton et al. 2004, 2006; Wittkowski et
al. 2007; Amiri et al. 2012; Fig.~\ref{fig:txcam}). 
Intriguing jet-like features are seen in some cases (Cotton et
al. 2006; Amiri et al. 2012), although their origin has remained a puzzle, as they cannot be
interpreted as simple outward accelerations. Temporal monitoring and
proper motions measurements are needed to establish  the true nature of these features.

\vspace{-0.4cm}
\paragraph{Variability}
SiO and H$_{2}$O masers in evolved stars are highly time variable (e.g., Pardo et
al. 2004; Kim et al. 2014).
The availability of the VLBA as a {\em dedicated} VLBI instrument has
thus been crucial for enabling the
regular monitoring of stellar masers with high spatial resolution. One
spectacular example is the 78-epoch ``movie'' of the SiO masers
in TX~Cam over nearly 5 years (Gonidakis et al. 2013;
Fig.~\ref{fig:txcam}).
These data indicate that a shock with velocity $\sim$7~\kms\ is created
during each stellar pulsation cycle that in turn affects the
intensity and distribution of the masers. Further, the velocity
structure suggests a bipolar geometry, contrary to the spherically symmetric outflows that are
traditionally assumed for AGB stars (see H\"ofner \& Olofsson 2018).
However, as we presently lack similar time-lapse data for other AGB stars, it is
impossible to draw general conclusions, and firm links between different
components of the atmospheric physics (shocks, pulsation, convection)
and the observed maser behaviors
are not yet established. Variability studies over a
large number of objects are needed to establish the connections between the AGB
atmosphere and the mass loss process.

%%%%%%%%%%%%%%%%%%%%%%%%%%%%%%%%%%%%%%%%
\begin{figure}[!t]
\vspace{-0.1cm}
\centering
\scalebox{0.4}{\rotatebox{0}{\includegraphics{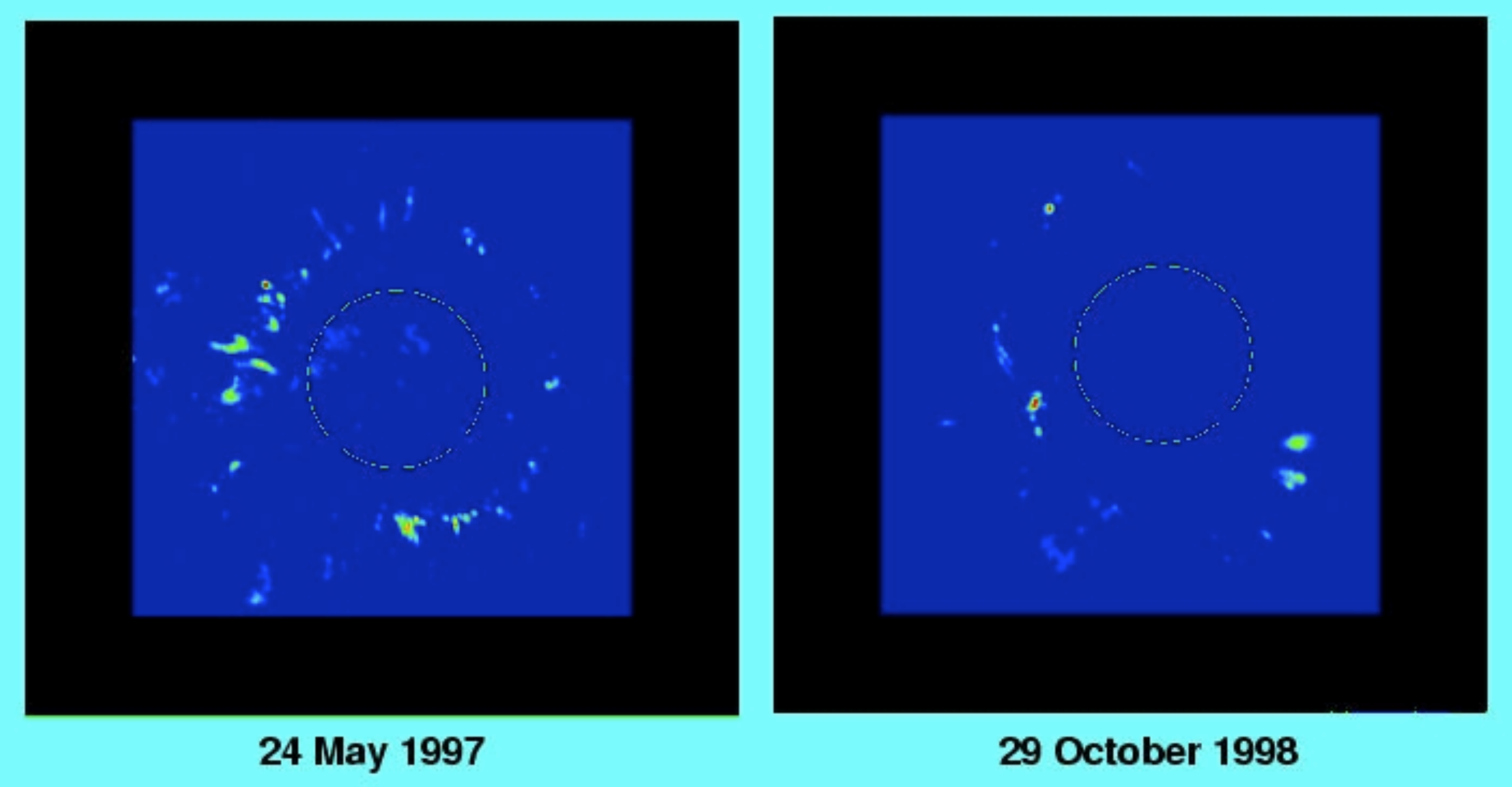}}}

\caption{{\small VLBA images of the $\lambda$7~mm SiO $v$=1, $J$=1-0 masers in
    TX~Cam during two different epochs. Credit: P. J. Diamond and
    A. J. Kemball (see also Diamond \& Kemball 2003; Gonidakis et al. 2013). }
\protect\label{fig:txcam}}
\vspace{-0.4cm}
\end{figure}
%%%%%%%%%%%%%%%%%%%%%%%%%%%%%%%%%%%%%%%

\vspace{-0.4cm}
\paragraph{Magnetic fields}
Full polarization measurements of SiO masers offer
a powerful means of constraining the little-understood role
of magnetic fields in AGB mass loss (e.g., Vlemmings 2018) and provide a vital link between
the ``surface'' magnetic field measured through infrared lines (L\`ebre et
al. 2014) 
with ``circumstellar'' magnetic fields measured further out
via H$_{2}$O and OH lines. Using the VLBA, Amiri et
al. (2012) obtained full-polarization maps of the
SiO masers in the OH/IR AGB star  OH~44.8-2.3 and discovered that they
are significantly linear polarized ($\sim$100\%), underscoring an
important role for magnetic fields in the outer atmosphere and
circumstellar environment (Fig.~\ref{fig:amiri}). The polarization vectors also seem
to indicate a dipolar magnetic field morphology, although the relationship between the
B-field geometry and the stellar outflow cannot yet be firmly
established. 
An improved
understanding of these results requires higher signal-to-noise ratio
observations, along with similarly detailed studies for
other AGB stars.

%%%%%%%%%%%%%%%%%%%%%%%%%%%%%%%%%%%%%%%%
\begin{figure}[!t]
\vspace{-0.1cm}
\centering
%\scalebox{0.3}{\rotatebox{0}{\includegraphics{Amiri_mom0.pdf}}}
\scalebox{0.4}{\rotatebox{0}{\vspace{-1.0cm}\includegraphics{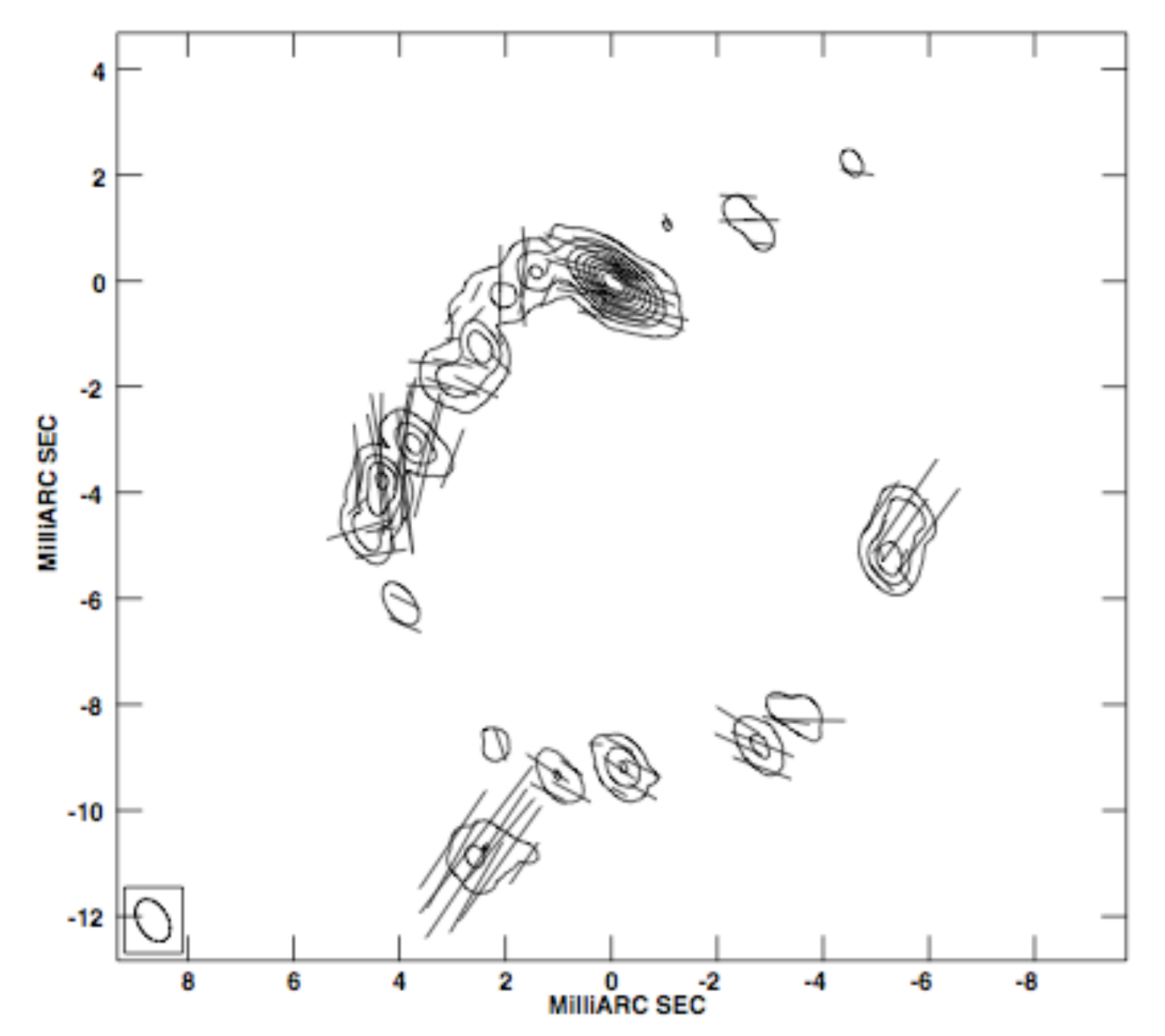}}}

\caption{{\small  Contour map of the SiO $v$=1, $J$=1-0  maser
    emission in the OH/IR star OH~44.8-2.3 obtained with the
    VLBA, overplotted with linear
   polarization vectors. Vector
    length is proportional to linearly polarized intensity (1 mas =
    1.25 Jy beam$^{-1}$) and position angle corresponds to the EVPA.  From Amiri et al. (2012). }
\protect\label{fig:amiri}}
\vspace{-0.5cm}
\end{figure}
%%%%%%%%%%%%%%%%%%%%%%%%%%%%%%%%%%%%%%%

\vspace{-0.3cm}
\paragraph{Multi-transition observations}
Multiple transitions and isotopologues of SiO and H$_{2}$O emit in the cm, mm, and
sub-mm (Alcolea 2004; Humphreys 2007; Gray 2012).  Because these various transitions require
different conditions to excite, spatially resolved
observations of multiple maser lines within a single star permit
measurements of
density and temperature within different regions of the atmosphere,
the propagation of shocks, and the transfer of material between
layers of the star (e.g.,
Humphreys et al. 2002; Gray et al. 2009, 2016). In particular, {\em contemporaneous}
observations of multiple lines and comparisons of their
properties and evolution with
those of the optical and radio photospheres offer 
potent diagnostics of the atmospheric physics. However, several key
lines emit outside the frequency coverage of the VLBA
(i.e., $\nu_{0}>$90~GHz) or else  fall below its brightness temperature
limits for line emission [e.g., $T_{B}\sim10^{8}$~K within a 31~kHz
spectral channel ($\sim$0.2~\kms\ for $\nu$=43~GHz) 
during a 6-hour integration].

Recently, a novel
optics system was installed on the Korean VLBI Network (KVN), enabling
simultaneous observations of four bands spanning 21--142~GHz (Han et
al. 2008). 
The promise of this set-up for observing stellar masers has already been
demonstrated (Cho et al. 2017; Yoon et al. 2018). However, the KVN lacks
the long baselines needed to  resolve the true sizes of maser emitting
gas clumps
and to gauge the fraction of emission emitted on various
spatial scales.  The longer baselines of the VLBA have
the needed resolution, but the limited instantaneous
bandwidth largely precludes the contemporaneous observations of
multiple lines. A consequence is persistent ambiguity in the astrometric
registration between different transitions that significantly
complicate the
interpretation of maser data
(e.g., Phillips et al. 2003; Desmurs et al. 2014; Issaoun et al. 2017; Fig.~\ref{fig:masers}).

%%%%%%%%%%%%%%%%%%%%%%%%%%%%%%%%%%%%%%%%%%%
 \begin{figure}[h]
%   \vspace{-1.cm}
   \hspace{0.75cm}
\centering
     \rotatebox{0}{
\resizebox{!}{5.5cm}
{\includegraphics{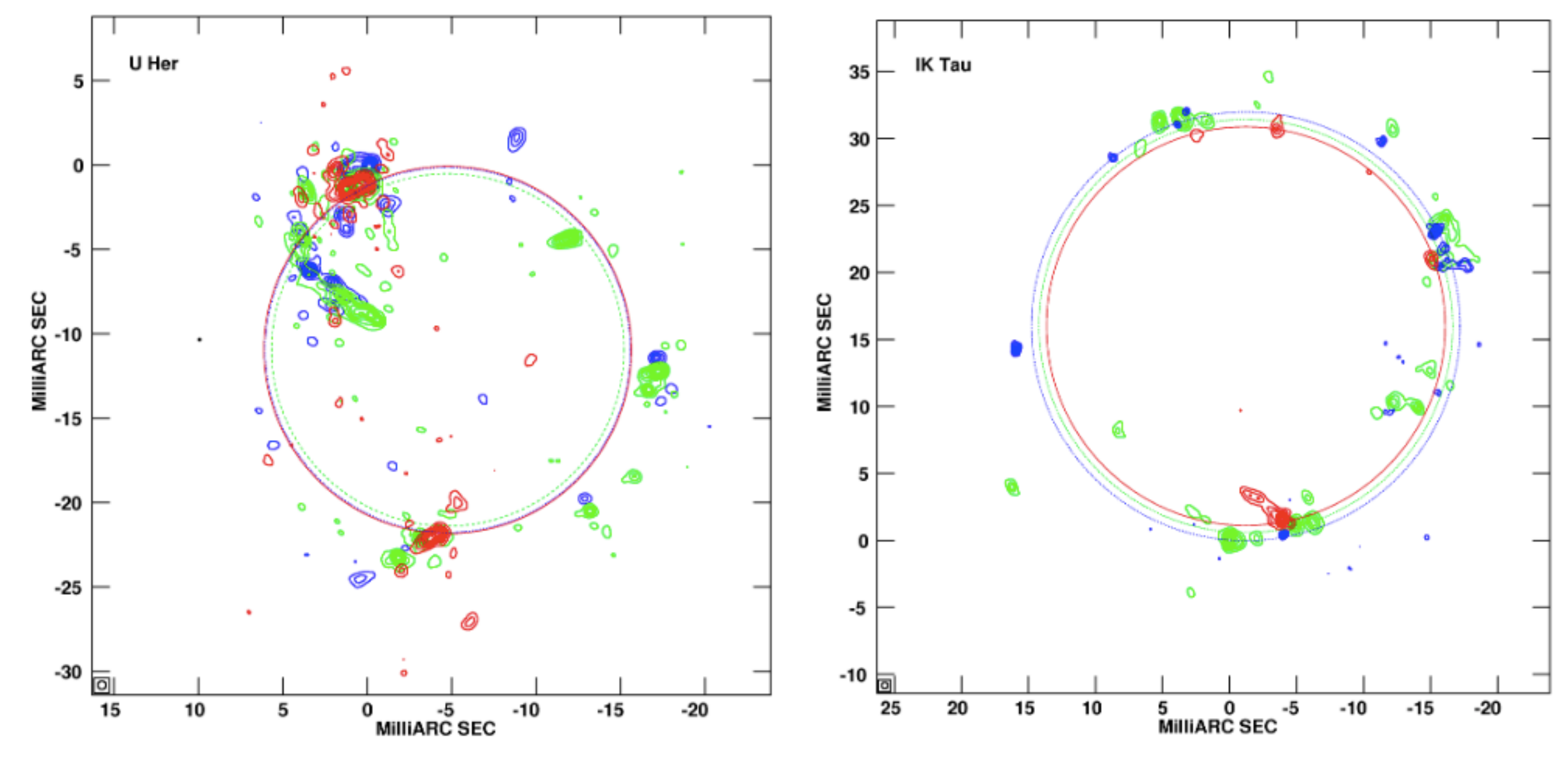}
     }}
         \caption{\small VLBA maps of SiO $v$ = 1
           (blue), $v$=2
           (green), and $v$=3 (red), \juc\ maser emission in
U~Her and IK~Tau (Desmurs et al. 2014). The relative
astrometry of the different  lines is currently highly
uncertain, limiting our ability to quantitatively constrain models
of the maser pumping and  atmospheric physics. }
    \label{fig:masers}
\vspace{-0.5cm}
\end{figure}
%%%%%%%%%%%%%%%%%%%%%%%%%%%%%%%%%%%%%%%%%%%%%%%%%%%%%%%

\section{Goals for the Next Decade: Requirements and Recommendations}

\vspace{-0.3cm}
Technological innovations during the next decade promise
major leaps in our ability to exploit VLBI studies of masers as a tool for
understanding stellar evolution and mass loss. 

\vspace{-0.4cm}
\paragraph{Goal: documenting temporal changes}
AGB star atmospheres are highly dynamic, making inferences gleaned from observations at only a
single observing epoch severely limiting---and potentially
misleading. The use of VLBI to obtain high time- and spatial-resolution ``movies'' of masers in
a sample of nearby ($d\lsim$1~kpc) AGB stars spanning a range in properties would supply vital
insights into the physics of AGB
outflows. For example, such observations would enable the measurement of
shock velocities (e.g., Gonidakis et al. 2013) which are critical to constraining the role
of pulsation in AGB mass loss (Reid \& Menten 1997; Gray et al. 2009) and
for explaining the possible existence of gas at
chromospheric temperatures (e.g., Luttermoser 1988; 
Vlemmings et al. 2017).  Such measurements can be compared with
independent assessments gleaned from the variability of
radio photosphere light curves (see Reid \& Menten 1997; Reid \&
  Goldston 2002). Full polarization observations would simultaneously enable
  constraints on the magnetic field strength and geometry (e.g.,
  Amiri et al. 2012).\\
{\bf Requirements/recommendations:} To enable full polarization 
monitoring of masers over both short and long cadences, 
it is crucial for the US community to maintain a dedicated VLBI
array with improved sensitivity. No other VLBI facility in the world has this capability in combination
with the $\sim10^{4}$~km baselines needed to provide
the angular resolution to fully
resolve the structure  and motions of individual maser clumps in stellar atmospheres. 

\vspace{-0.4cm}
\paragraph{Goal: multi-frequency line mapping} Building a complete understanding of the physical conditions,
chemistry, and gas motions within the atmospheres and envelopes of AGB
stars benefits from the ability to detect and
simultaneously observe a wide range of maser
transitions, including relatively
weak ($T_{B}\sim10^{4}$~K) and little explored lines such as
SiO $v$=0   (Bolboltz \& Claussen  2004); $^{29}$SiO $v$=0,1, $^{28}$SiO $v$=2,
\jdu, $^{28}$SiO $v$=3  (Soria-Ruiz et al. 2005, Desmurs et
al. 2014), and  HCN (unique to carbon stars; e.g., Izumiura et al. 1995; Menten et al. 2018).\\
{\bf Requirements/recommendations:} Enabling contemporaneous VLBI measurements of multiple
maser lines requires upgrading
VLBA stations to wider
instantaneous bandwidths ($\gsim$8~GHz) and expanded frequency
coverage (to $\nu\gsim$100~GHz). While wider bandwidths do not increase
spectral line sensitivity, they improve measurements by expanding
the available high-frequency calibration sources. 
Parallel improvements in line
sensitivity can be achieved through: (1)
inclusion in VLBI arrays additional large apertures such as phased
ALMA ($0.8\lsim \lambda\lsim 7$~mm;
Matthews et al. 2018), the phased VLA ($0.7\lsim \lambda\lsim 1.3$~cm),
and the Robert C. Byrd Green Bank Telescope ($0.3\lsim \lambda\lsim
1.3$~cm); and (2) the addition of many
stations on intermediate baselines ($\sim$30-300~km) to bridge the spatial
scales and brightness temperature sensitivity of current VLBI arrays and connected element
interferometers (see Kameno et al. 2013; Selina et al. 2018). A
continuum brightness temperature
sensitivity of $<10^{3}$~K on the intermediate baselines would enable
simultaneous detection and astrometric registration of the stellar
continuum and the maser emission
with exquisite precision ($\ll R_{\star}$), providing valuable new insights into the transport
of matter and energy in AGB star atmospheres and into the pumping
mechanism for masers (e.g., Gray et al. 2009,
2016; Desmurs et al. 2014).

\end{justify}

\pagebreak
\raggedright
\textbf{References}

\begin{hangparas}{.25in}{1}
Alcolea, J., Bujarrabal, V., \& Gallego, J. D. 1989, A\&A, 211, 187

Amiri, N., Vlemmings, W. H. T., Kemball, A. J., \& van Langevelde,
H. J. 2012, A\&A, 538, A136

Blackman, E. G., Frank, A., \& Welch, C. 2001, ApJ, 546, 288

Boboltz, D.~A., \& Claussen, M.~J.\ 2004, ApJ, 608, 480

Bowen, G. H. 1988, ApJ, 329, 299

Cho, S.-H., Yun, Y., Kim, J., et al. 2018, IAU Symp. 336, ed. A. Tarchi,
M. J. Reid, \& P. Castangia, 359

Cotton, W. D., Mennesson, B., Diamond, P. J., et al. 2004, A\&A, 414, 275

Cotton, W. D., Vlemmings, W., Mennesson, B., et al. 2006, A\&A, 456,
339

Desmurs, J. F., Bujarrabal, V., Colomer, F., \& Alcolea, J. 2000,
A\&A, 360, 189

Diamond, P. J. \& Kemball, A. J. 2013, ApJ 599, 1372
	
Diamond, P. J., Kemball, A. J., Junor, W., Zensus, A., Benson, J., \&
Dhawan, V. 1994, ApJ, L61

Freytag, B., Liljegren, S., \& H\"ofner, S. 2017, A\&A, 600, A137

Gonidakis, I., Diamond, P. J., \& Kemball, A. J. 2013, MNRAS, 433,
3133

Gray, M. 2012, Maser Sources in Astrophysics, (Cambridge: Cambridge
University Press)

Gray, M. D., Baudry, A., Richards, A. M. S., Humphreys, E. M. L.,
Sobolev, A. M., \& Yates, J. A. 2016, MNRAS, 456, 374

Gray, M. D., Wittkowski, M., Scholz, M., Humphreys, E. M. L., Ohnaka,
K., \& Boboltz, D. 2009, MNRAS, 394, 51

Han S.-T., Lee, J.-W. Kand, J. et al. 2008, IJIMW, 29, 69

Harper, G. M. 2010, ApJ, 720, 1767

H\"ofner, S. 2011, ASPC, 445, 193

H\"ofner, S. \& Freytag, B. 2019, A\&A, in press (arXiv:1902.04074)

H\"ofner, S. \& Olofsson, H. 2018, A\&ARv, 26, 1

Humphreys, E. M. L. 2002, IAU Symp. 206, 266

Humphreys, E. M. L. 2007, IAU Symp. 242, 471

Humphreys, E. M. L., Gray, M. D., Yates, J. A., Field, D., Bowen,
G. H., \& Diamond, P. J. 2002, A\&A, 386, 256

Issaoun, S., Goddi, C., Matthews, L. D., et al. 2017, A\&A, 606, A126

Izumiura, H., Ukita, N., \& Tsuji, T. 1995, ApJ, 440, 728

Kameno, S. Nakai, N., \& Honma, M. 2013, New Trends in Radio Astronomy
in the ALMA Era, 
ASP Conference Series, 476, (San Francisco: ASP), 409

Karakas, A. I. 2014,  IAU Symp.
298, ed. S. Feltzing, G. Zhao, N. A. Walton, \&
P. A. Whitelock, (Cambridge: Cambridge University Press), 142

Kim, J., Cho, S.-H., \& Kim, S. J. 2014, AJ, 147, 22

Kwok, S. 1975, ApJ, 198, 583

L\`ebre, A., Auri\`ere, M., Fabas, N., Gillet, D., Herpin, F.,
Konstantinova-Antova, R., \& Petit, P. 2014, A\&A, 561, 85

Leitner, S. N. \& Kravtsov, A. V. 2011, ApJ, 734, 48

Lim, J., Carilli, C. L., White, S. M., Beasley, A. J., \& Marson,
R. G. 1998, Nature, 392, 575

Luttermoser, D. G. 1988, PASP, 100, 1587

Marengo, M.  2009, PASA, 26, 365

Matthews, L. D., Crew, G. B., Doeleman, S. S., et al. 2018, PASP, 130, 5002

Matthews, L. D., Greenhill, L. J., Goddi, C., Chandler, C. J., Humphreys,
E. M. L., \& Kunz, M. W. 2010, ApJ, 708, 80

Melnick, J. \& De Propris, R. 2013, MNRAS, 431, 2034

Menten, K. M., Wyrowski, F., Keller, D., \& Kami\'nski, T. 2018, A\&A,
613, 49

Neilson, H. R. 2014, IAU Symp. 301, 205

O'Gorman, E., Harper, G. M., Guinan, E. F., Richards,
  A. M. S., Vlemmings, W., \& Wasatonic, R. 2015, A\&A, 580, A101

O'Gorman, E., Kervella, P., Harper, G. M., Richards, A. M. S., 
Decin, L., Montarg\` es, M., \& McDonald, I. 2017, A\&A, 602, L10

Pardo, J. R., Alcolea, J., Bujarrabal, V., Colomer, F., del Romero,
A., \&  de Vicente, P. 2004, A\&A, 424, 145

Phillips, R. B., Straughn, A. H., Doeleman, S. S., \& Lonsdale,
C. J. 2003, ApJ, 588, 105

Reid, M. J. \& Goldston, J. E. 2002, ApJ, 568, 931

Reid, M. J. \& Menten, K. M. 1997, ApJ, 476, 327

Schr\"oder, K.-P. \& Sedlmayr, E. 2010, A\&A, 366, 913

Selina, R. J., Murphy, E., J., McKinnon, M., et al. 2018, in Science with a Next Generation Very Large Array, ASP
Monograph 7, ed. E. J. Murphy (San Francisco: ASP), 15

Soria-Ruiz, R., Alcolea, J., Colomer, F., Bujarrabal, V., Desmurs,
J.-F., Marvel, K. B., \& Diamond, P. J. 2004, A\&A, 426, 131

Snyder, L. E. \& Buhl, D. 1974, ApJ, 189, L31

Tosi, M. 2007, ASPC, 368, 353

Van Eck, S., Goriely, S., Jorissen, A., \& Plez, B. 2001, Nature, 412,
793

Vlemmings, W. H. T. 2018, CoSka, 48, 187

Vlemmings, W., Khouri, T., O'Gorman, E., et al. 2017,  Nature Ast, 1, 848

Wilson, W. J. \& Barrett, A. H. 1968, AJ, 73, 209

Wittkowski, M., Boboltz, D. A., Ohnaka, K., Driebe, T., \& Scholz,
M. 2007, A\&A, 470, 191

Woitke, P. 2006, A\&A, 460, 9

Yoon, S.-C. \& Cantiello, M. 2010, ApJ, 717, 62

Yoon, D.-H., Cho, S.-H. Yun, Y., et al. 2018, Nature Comm., 9, 2534
\end{hangparas}
\end{document}